\documentclass[twocolumn,showpacs,superscriptaddress,10pt]{revtex4-1}
\usepackage{amsmath,amssymb,graphicx}
 \pdfoutput=1

\usepackage{upgreek}
\usepackage{hyperref}

\begin{document}

\title{Radiocarbon Dioxide detection based on Cavity Ring-Down Spectroscopy and a Quantum Cascade Laser}

\author{G.~Genoud}
\email{guillaume.genoud@vtt.fi}
\affiliation{VTT Technical Research Centre of Finland Ltd, Centre for Metrology MIKES, P.O. Box 1000, FI-02044 VTT, Finland}
\author{M.~Vainio}
\affiliation{VTT Technical Research Centre of Finland Ltd, Centre for Metrology MIKES, P.O. Box 1000, FI-02044 VTT, Finland}
\affiliation{Laboratory of Physical Chemistry, Department of Chemistry, 
P.O. Box 55, FI-00014 University of Helsinki, Finland}
\author{H.~Phillips}
\affiliation{National Physical Laboratory, Hampton Road, Teddington, Middlesex, TW11 0LW, UK}
\author{J.~Dean}
\affiliation{National Physical Laboratory, Hampton Road, Teddington, Middlesex, TW11 0LW, UK}
\author{M.~Merimaa}
\affiliation{VTT Technical Research Centre of Finland Ltd, Centre for Metrology MIKES, P.O. Box 1000, FI-02044 VTT, Finland}

\begin{abstract}

Monitoring of radiocarbon ($^{14}$C) in carbon dioxide is demonstrated using mid-infrared spectroscopy and a quantum cascade laser.
The measurement is based on cavity ring-down spectroscopy, and a high sensitivity is achieved with a simple setup. The instrument was tested using a standardised sample containing elevated levels of radiocarbon. Radiocarbon dioxide could be detected from samples with an isotopic ratio $^{14}$C/C as low as 50 parts-per-trillion, corresponding to an activity of 5 kBq/m$^3$ in pure CO$_2$, or 2 Bq/m$^3$ in air after extraction of the CO$_2$ from an air sample.
The instrument is simple, compact and robust, making it the ideal tool for on-site measurements. It is aimed for monitoring of radioactive gaseous emissions in nuclear power environment, during the operation and decommissioning of nuclear power plants. Its high sensitivity also makes it the ideal tool for the detection of leaks in radioactive waste repositories.

This paper was published in \textit{Optics Letters} and is made available as an electronic reprint with the permission of OSA. The paper can be found at the following URL on the OSA website: \url{http://www.opticsinfobase.org/ol/abstract.cfm?uri=ol-40-7-1342}. Systematic or multiple reproduction or distribution to multiple locations via electronic or other means is prohibited and is subject to penalties under law.
\end{abstract}


\maketitle

Detection of trace amount of gases is important for many scientific and industrial applications, such as air quality and industrial emissions monitoring, or detection of hazardous substances. In this context, optical spectroscopy has shown to be an important tool by providing sensitive and fast measurements capabilities.
The ability of laser spectroscopy to measure isotopic ratios of molecules also led to many applications.
The radiocarbon isotope ($^{14}$C) has a natural abundance of $^{14}$C/C=1.2 parts-per-trillion (ppt), but can be found at elevated levels in nuclear facilities. It is present in all parts of nuclear power plants  and most of it has potential for gas-phase release, mostly in the form of carbon dioxide \cite{Yim2006}. Airborne radioactive releases are thus dominated by $^{14}$CO$_2$, and are a major concern during operation and decommissioning of nuclear facilities. For instance, $^{14}$CO$_2$ is produced in waste repositories, where biodegradation of radioactive waste leads to $^{14}$CO$_2$ emissions with an activity concentration of the order of 1-100 Bq/ml, corresponding to $^{14}$C/C=10 ppb - 1 ppm \cite{Thorne}. Operational gaseous emissions take place inside nuclear facilities and through the stack air with a $^{14}$C activity concentration that has been measured to be 200 Bq/m$^3$ \cite{Stenstrom1995}. 
In the future, the amount of waste to be treated and monitored will increase, and so will potential emissions. 
Sensitive measurements of  $^{14}$CO$_2$ are thus required to detect the smallest leaks of contaminated gases, and ensure a good monitoring of releases from nuclear facilities and waste repositories.
In this Letter we report on the development of a mid-infrared spectrometer designed to monitor elevated level of $^{14}$CO$_2$ in nuclear facilities. Thanks to the use of a quantum cascade laser (QCL) as light source and a simple design, the instrument is compact, ideal for field measurement. To our knowledge it is the first time that radiocarbon dioxide is detected using a QCL.

Radiocarbon is the ideal tracer to discriminate between emissions of fossil or biogenic origin. It has been used for biofraction measurement \cite{Haemaelaeinen2007,Mohn2008}, microdosing studies \cite{Lappin2004} and carbon dating. Detection of $^{14}$C with optical techniques has recently gained interest as  technology for mid-infrared instrumentation has improved. The state of the art for its measurement is accelerator mass spectrometry (AMS), which can achieved extremely high sensitivity. Liquid scintillation counting is also used. However, these methods have neither on-line nor on-site capabilities, and usually require complex sample preparation. The measurement procedure with AMS can, for instance, take several days. There is therefore a need for a technique with in-situ measurement capabilities.

Detection of $^{14}$CO$_2$ using diode laser spectroscopy has been studied in the 1980’s \cite{Wahlen1977,Labrie1981}, but with the technology of that time the projected detection sensitivity was relatively low. Cavity ring-down spectroscopy (CRDS) was later proposed for nuclear power monitoring \cite{Tomita2008}.
Recently, amazing sensitivity was achieved using an advanced spectroscopic technique called saturated cavity ring-down \cite{Galli2011}. However, this technique requires a large and complex light source, not suitable for in-situ measurements.
The novelty of this work is the use of a continuous-wave QCL as a light source for the detection of $^{14}$CO$_2$. High resolution can be achieved using such QCLs that have a linewidth of a few tens of MHz \cite{Bartalini2011}. They are relatively inexpensive compared to other mid-infrared laser sources, and together with their small size, they allow for the development of a compact and low cost instrument with a great potential for on-site measurements. To reach the necessary sensitivity, continuous-wave CRDS is used \cite{Romanini1997}.

Spectral data for $^{14}$CO$_2$ are available in a few publications \cite{Wahlen1977,Labrie1981,Galli2011a}, and the strongest absorption band for the detection of $^{14}$CO$_2$ corresponds to the fundamental asymmetric stretching vibration band $\nu_3$. Due to large isotopic shifts in CO$_2$, interferences from other isotopes are minimal.
In particular, the P(20) line at 4.527~$\upmu$m (2209.1~cm$^{-1}$) has been shown to be suitable for the detection of $^{14}$CO$_2$ using laser spectroscopy \cite{Wahlen1977,Labrie1981,Galli2011}.
The absorption coefficient, $\kappa$, is expressed in terms of cavity exponential decay time $\tau(\nu)$, also called ring-down time, via $\kappa(\nu)=(\tau_0-\tau(\nu))/(c \tau(\nu) \tau_0)$, where $\tau_0$ is the ring-down time in vacuum, $\nu$ the wavenumber, and $c$ the speed of light \cite{Berden2000}.

The measurement cell is formed by a 40 cm long cavity with mirrors of reflectivity $\sim~99.98\%$ which yields to a typical vacuum ring-down time of $\sim~5.5$~$\upmu$s.
Using two spherical mirrors, the mode of the laser is matched to the mode of the cavity. The laser, a distributed feedback QCL from Hamamatsu (L12004-2209H-C), delivers up to 15 mW, of which 3 mW reaches the cavity. The light exiting the cavity is then focused onto   an infrared photovoltaic detector (PVI-2TE-5-1 by VIGO).
A metering valve allows for a controlled injection of the gas into the cavity, and the pressure is measured with a capacitance manometer.
The temperature of the cell is not actively stabilised but is the same as the laboratory temperature which is stabilised and monitored. 
The footprint of the instrument is only $90 \times 60$ cm$^2$, and can easily be reduced by optimising the design of the setup. 
Light is coupled into the cavity by applying a linear ramp to the laser current at typically 10~Hz rate, resulting in a linear scan of the wavenumber (a one inch long Germanium etalon provides a more precise calibration). The ramp covers several longitudinal modes of the cavity. At the same time, the cavity length is modulated at a slower rate to obtain frequency points between the fixed points determined by the free spectral range of the cavity.  
The ring-down events are produced by applying a rapid positive step to the laser current, which causes the laser wavelength to suddenly change and coupling into the cavity to stop. As the bandwidth of the current driver (QCL1000 from Wavelength Electronics) is limited to 2~MHz, the step is not steep enough to immediately stop coupling, and initially light continues to enter the cavity. The first 2~$\upmu$s of the ring-down event are therefore ignored. The signals recorded by the detector are digitized by a 14 bits, 100 MHz acquisition card. The data is then processed with LabView and a fitting algorithm determines the ring-down time using an exponential decay  model. Finally, the absorption spectrum is inferred from the ring-down times recorded as function of the laser wavenumber, which is deduced from the measured QCL current when the trigger occurs.

Optical feedback was found to affect the laser, due to light transmitted from the cavity back to the QCL. This feedback considerably narrows the linewidth of the laser, which effectively becomes the linewidth of the cavity \cite{Laurent1989}, but adds uncertainty in the lasing frequency of the laser, which does not correspond to the frequency set by the current driver any more. A Faraday isolator (from Electro-Optics Technology) was thus placed after the QCL, efficiently preventing the feedback from the cavity to reach the laser.
The sensitivity was also limited by interferences, due to reflections from the different optical surfaces back into the cavity \cite{Fox2002}. To minimize these reflections, spherical mirrors are used as mode matching optics instead of lenses. The remaining reflections from the optical isolator and the detector produce interference fringes with a frequency of the same order as the absorption line width. The lengths between the cavity and these optical surfaces are thus modulated using piezoelectric translation-mounted mirrors. This way the phase of the back reflection is randomised, and the interference fringes are removed after averaging~\cite{Silver1988}.

A standardised gas sample with elevated levels of $^{14}$C was produced at the National Physical Laboratory (NPL). $^{14}$CO$_2$ was generated from a solution of NaH$^{14}$CO$_3$. The initial sample had an activity of 3608 Bq/ml, and was diluted in two steps by adding a known amount of inactive CO$_2$ to first 10 Bq/ml and then 0.1 Bq/ml. The mass of gas transferred was determined gravimetrically. The gas dilution factor and thus the activity was confirmed by sampling and trapping of gas followed by liquid scintillation counting of the trapping medium. The final activity concentration was determined to be 0.096 Bq/ml $\pm 0.008$ ($1\upsigma$) corresponding to a concentration of R=$^{14}$C/C = 1.007 $\pm 0.085$~ppb.

\begin{figure}[b]
\centering
\includegraphics[width=1\columnwidth]{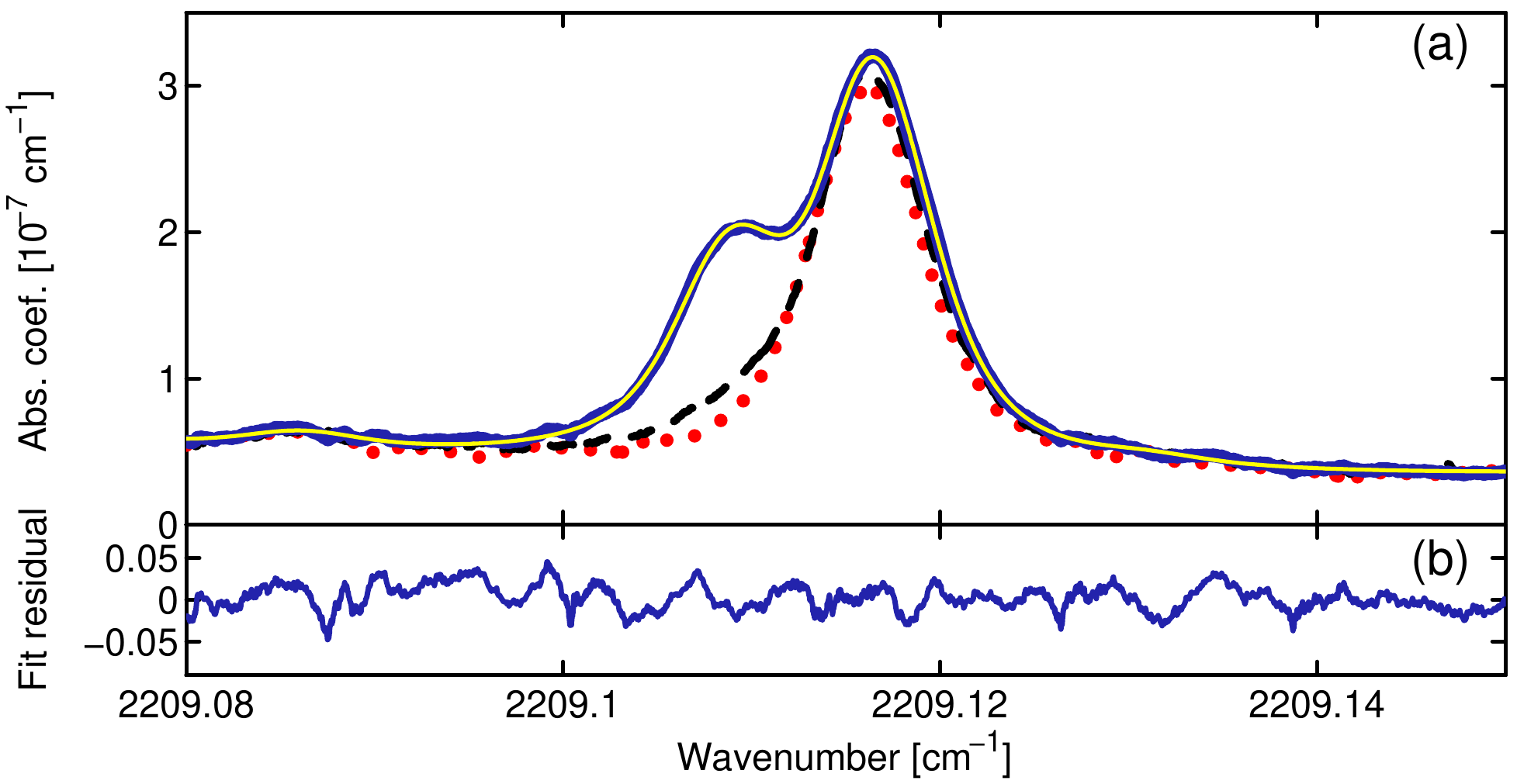}
\caption{Absorption spectra of pure CO$_2$ with different isotopic ratios at a pressure of 20 mbar in (a). The sample from NPL with 1~ppb of $^{14}$C (solid blue line) was compared with an inactive sample (dotted red line). An additional absorption peak due to $^{14}$CO$_2$ is visible. The black dashed line shows a spectrum with 110 ppt of $^{14}$C.  For clarity the spectra were smoothed using a moving average filter. In yellow, a fit of the blue line is shown with its residual in (b).}
\label{c14_specs}
\end{figure}

\begin{figure}[t]
\centering
\includegraphics[width=1\columnwidth]{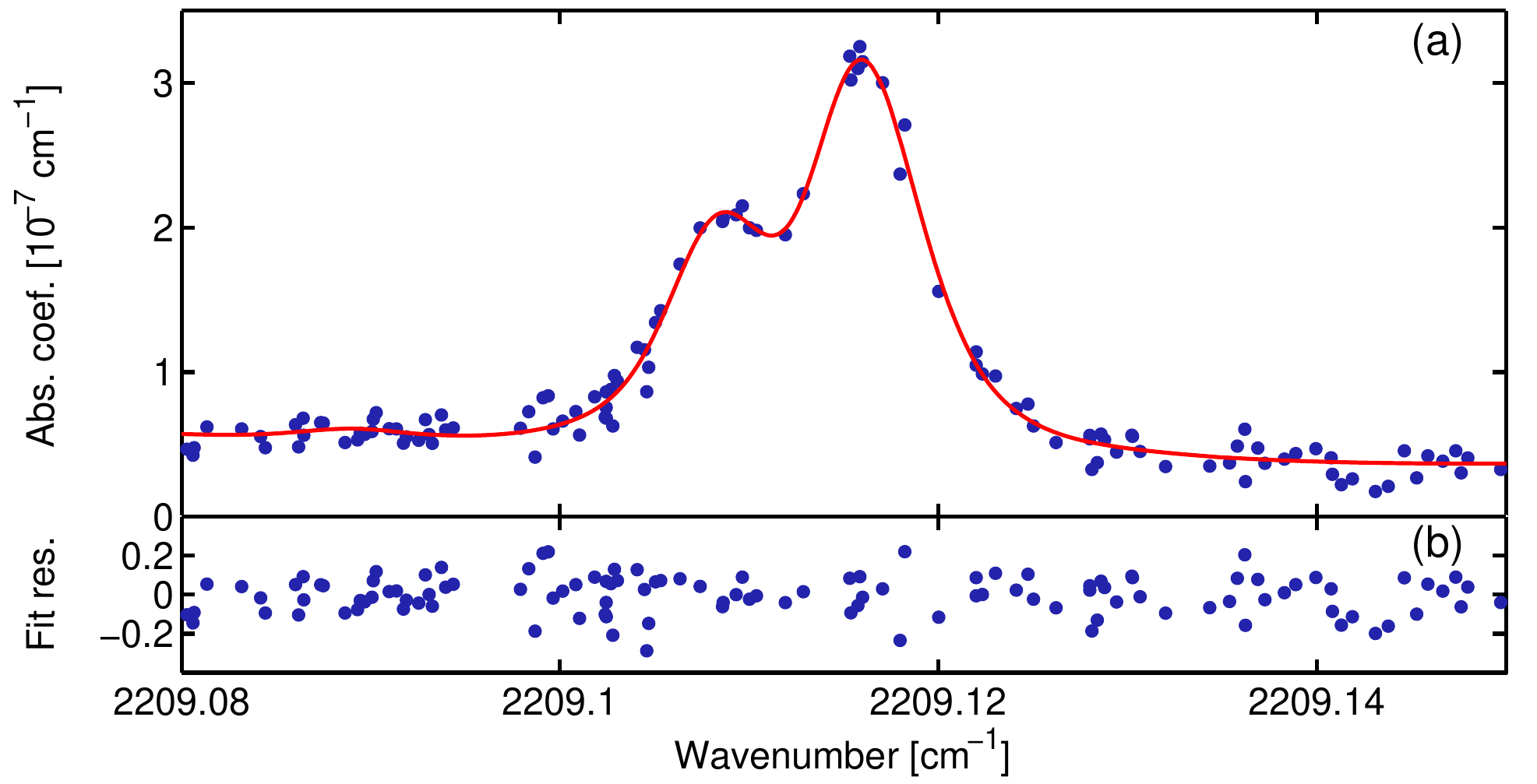}
\caption{Absorption spectra recorded in 10 sec at a pressure of 20 mbar, which is sufficient to acquire an entire spectrum and determine the radiocarbon content. In red, a fit of the data points is shown and its residual can be seen in~(b).}
\label{spec_10s}
\end{figure}

Absorption spectra were recorded in the wavelength region of interest. Each spectrum was recorded over a period of 5 min and consisted of about 5000 ring-down events. The absorption coefficient was computed with a vacuum ring-down time, $\tau_0=5.45$~$\upmu$s, determined from the baseline. A reference spectrum of inactive pure CO$_2$ was first analysed at a pressure of 20 mbar and compared with the standardised sample prepared by NPL, as shown in Fig. \ref{c14_specs} (a). One observes an additional absorption peak appearing. As expected, this is the P(20) peak of $^{14}$CO$_2$ at 2209.11~cm$^{-1}$. The other visible peaks correspond to different stable isotopes of CO$_2$. In particular, the strong peak on the right of the $^{14}$CO$_2$ peak consists of two closely spaced vibrational levels of the $^{16}$O$^{13}$C$^{16}$O isotope. Fig. \ref{spec_10s} shows an absorption spectrum recorded over only 10~sec, where the $^{14}$CO$_2$ peak is clearly visible. The instrument can detect $^{14}$CO$_2$ at a fast rate, showing real on-line measurement capabilities.

The absorption spectra were fitted by a sum of Voigt profiles using a non linear least square fitting routine. In Fig. \ref{c14_specs} (a) and Fig. \ref{spec_10s} (a), the obtained fits are plotted, with their residuals shown in (b). In order to correctly model the background, it was necessary to take into account absorption lines of other CO$_2$ isotopes situated close to the targeted absorption line of $^{14}$CO$_2$. Four main lines corresponding to other carbon dioxide isotopes were identified from the HITRAN database \cite{hitran}
 at 2208.95 ($^{16}$O$^{13}$C$^{16}$O), 2208.99 ($^{16}$O$^{13}$C$^{18}$O), 2209.01 ($^{16}$O$^{13}$C$^{16}$O) and 2209.37 ($^{16}$O$^{12}$C$^{16}$O). Together with some other weaker lines, a total of 12 lines were included in the fitting model.

The line strength of the P(20) transition is determined by: $S_0(T)=S k_B T/(R p)$, with $R$ the known concentration of $^{14}$C in the NPL sample, $S$ the line area of the P(20) line obtained from the fit, $p$ the measured pressure in the cavity, $k_B$ the Boltzmann constant, and $T$ the sample temperature, which was 294.5~K.
We obtain $S_0(T=294.5 K)=2.52\pm 0.26 \times10^{-18}$cm$^{-1}/$(molecule cm$^{-2}$), in good agreement with Whalen \textit{et al.} \cite{Wahlen1977}, where the line strength was measured experimentally to be $S_0=2.5\times10^{-18}$cm$^{-1}/$(molecule cm$^{-2}$).
To assess the performance of the measurement at fast rates, 
the activity of the 10 sec scan shown in Fig. \ref{spec_10s} was calculated. Using the line strength determined earlier it was found to be 0.105 Bq/ml.  A larger measurement uncertainty was expected, as the residual of the fit is worst than with longer averaging time.

\begin{figure}[b]
\centering
\includegraphics[width=0.75\columnwidth]{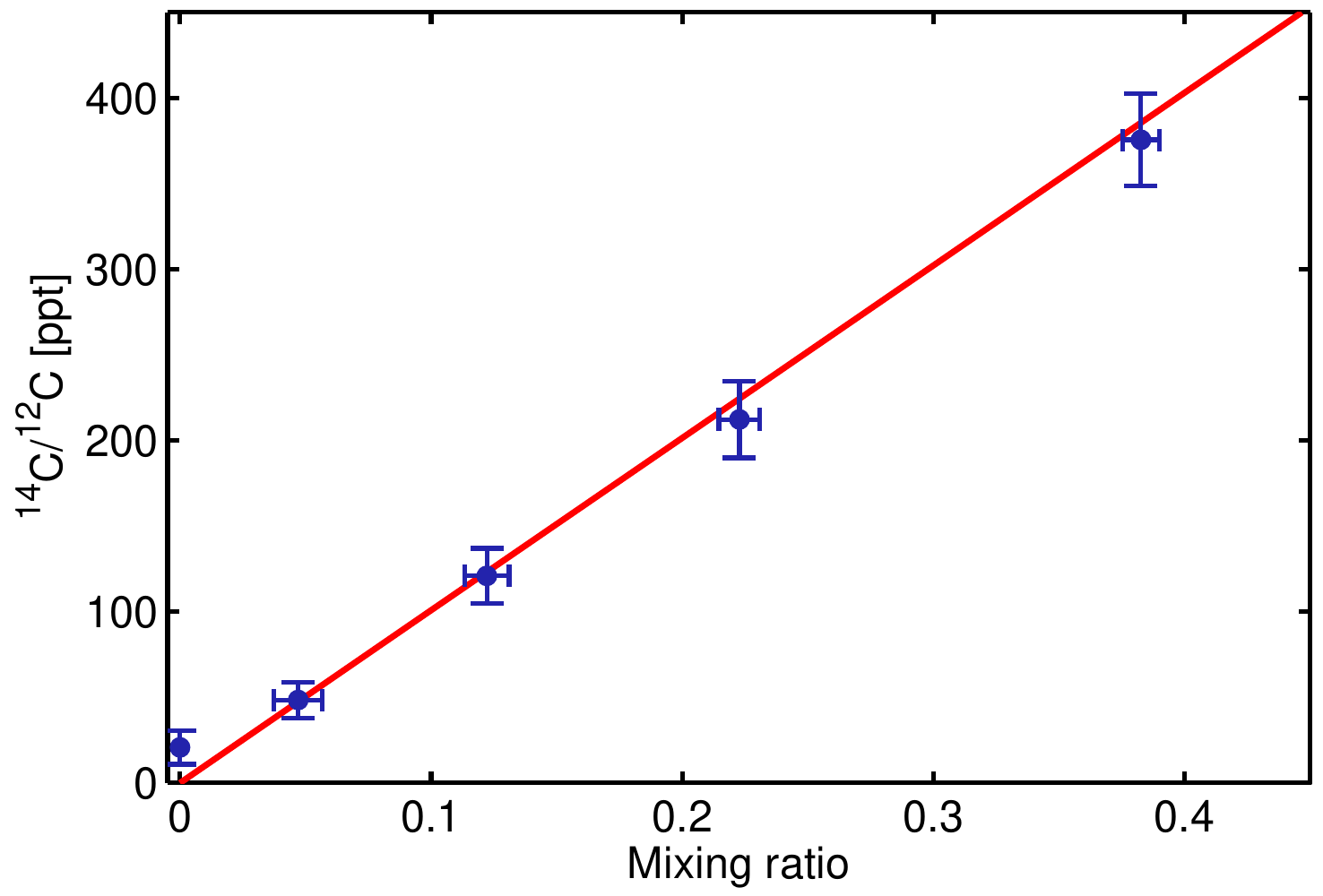}
\caption{Isotopic ratio calculated from the measured line areas as function of mixing ratio between the NPL sample with elevated levels of $^{14}$C and inactive CO$_2$. The red line is the expected behaviour.}

\label{concentration}
\end{figure}

The behaviour of the instrument with samples containing lower concentrations of $^{14}$C was also investigated.
The initial sample was diluted by filling the measurement cell to a known pressure using the NPL sample and then completing it with pure, inactive CO$_2$ up to about 20 mbar. The black dashed line in Fig. \ref{c14_specs} (a) shows such a spectrum with 22\% of the NPL sample and 78\% of CO$_2$, i.e. 110 ppt of $^{14}$C. While the strength of the $^{14}$CO$_2$ line has greatly decreased, it is still visible. 
Fig. \ref{concentration} shows the calculated isotopic ratio $R= {^{14}C}/C$ as function of mixing ratio. $R$ was determined for each mixture, by measuring the line areas of the $^{14}$CO$_2$ and $^{13}$CO$_2$ isotope absorption peaks, and normalising it using the ratio of the line areas of the NPL sample which has a known isotopic ratio. The data points follow the expected linear trend represented by the red line. The same calculation was carried out with a sample containing no $^{14}$C, corresponding to the data point at a mixing ratio of 0. It is seen that the calculated isotopic ratio is not 0, but 20~ppt. This is mainly due to the background consisting of absorption lines of other isotopes, which produces an uncertainty in the fit, thus limiting the sensitivity of the instrument. The detection limit is expected to increase significantly by cooling down the sample, as cooling efficiently reduces interferences from other isotopes caused by hot transitions \cite{Labrie1981}. However, for monitoring of radioactive emissions, this additional step is not necessary as radiocarbon levels would be relatively high. We estimate the current detection limit of our instrument to be 50~ppt or 5 kBq/m$^3$.
The 10 sec scan shown in Fig. \ref{spec_10s} yields to a noise equivalent detection sensitivity ($1\upsigma$) of $2.1\times 10^{-9}$ cm$^{-1}$Hz$^{-0.5}$.

\begin{figure}[t]
\centering
\includegraphics[width=0.75\columnwidth]{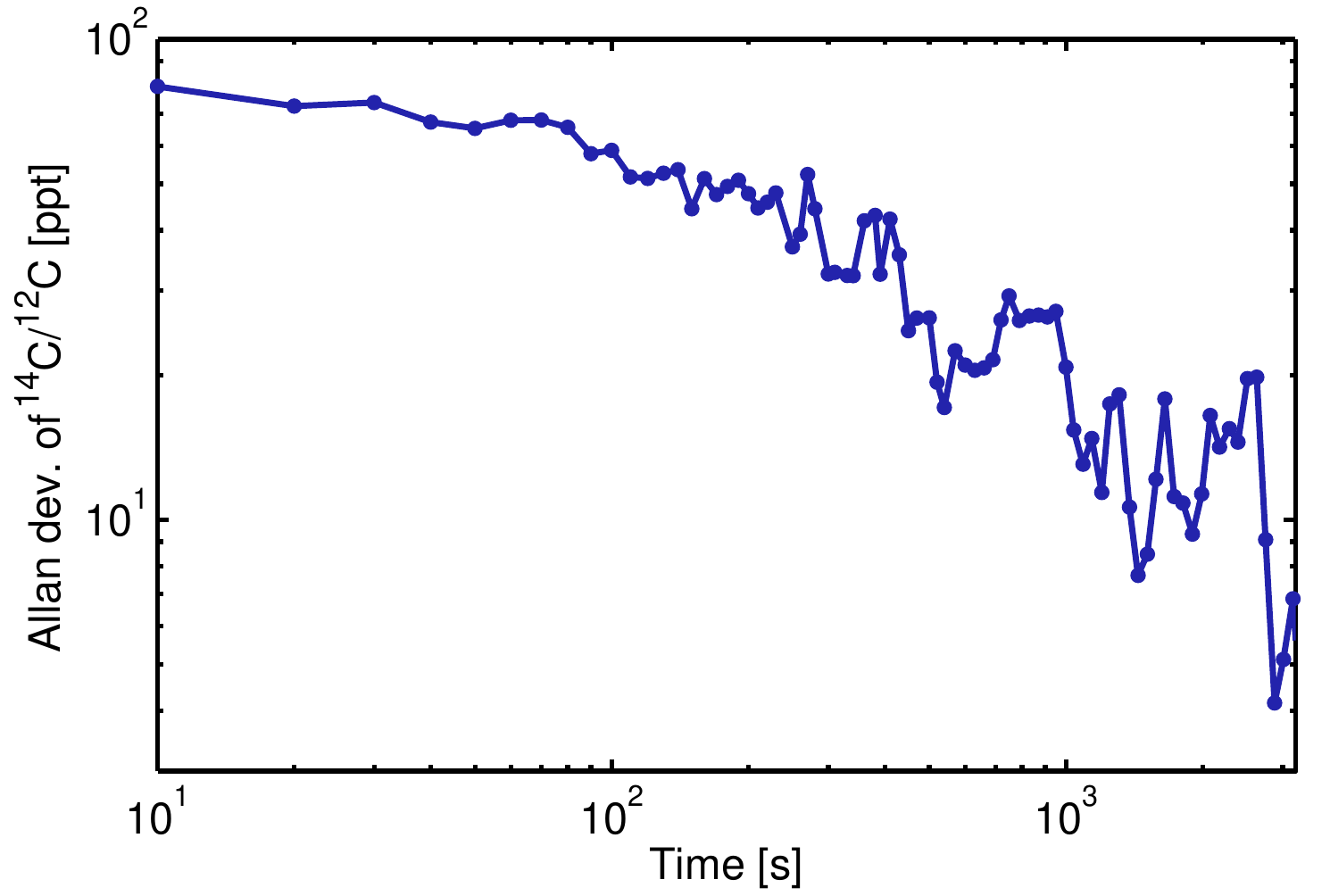}
\caption{Allan deviation plot of the isotope ratio ${^{14}C}/C$ of a scan with a sample containing 1 ppb of $^{14}$C. From the graph, the measurement resolution is estimated to be 10 ppt.}
\label{allan}
\end{figure}

Stability of the measurement was investigated using Allan deviation. A two hours long scan of the targeted absorption lines was sliced in 20 sec scans. For each of those scans, the isotopic ratio $R= {^{14}C}/C$ was calculated, and the Allan deviation computed as shown in Fig. \ref{allan}.
Due to the very stable laboratory environment of the metrology laboratory and the robust cavity arrangement, our measurement is very stable over several hours, explaining the fact that the Allan deviation does not level off during the observed time period.
This allows for long averaging times, and increased sensitivity. The estimated resolution is given by the minimum in the Allan deviation, 10~ppt. We observed drifts in the line position attributed to drifts in the laser wavelength ($\sim 0.0015$ cm$^{-1}$ over one hour). However, this does not affect the line area measurement as long as the acquisition time is shorter than the drift period. By recording spectra at fast rate, and then average the obtained line areas, the measurement becomes immune to drifts. This method was used to produce the data shown in Fig. \ref{concentration}.

In conclusion, radiocarbon dioxide at concentration levels relevant to the nuclear industry, was detected using laser spectroscopy and a quantum cascade laser. The instrument was tested with sample of pure CO$_2$ with an elevated level of $^{14}$C. In field conditions, samples are not pure CO$_2$, but air. CO$_2$ has to be first extracted from air using, for example, a cryogenic trap. Other compounds containing $^{14}$C such as methane could be analysed by burning them first into CO$_2$. 
The detection limit is estimated to be ${^{14}C}/C=50$ ppt. This value can be extrapolated to a measurement of CO$_2$ extracted from an air sample, where the CO$_2$ concentration is 400 ppm. 50 ppt corresponds then to an activity of 2 Bq/m$^3$ in air. The activity concentration in the stack air of nuclear facilities is $\sim200$ Bq/m$^3$,  and early detection of those radioactive emissions are possible with our instrument. In situations where high activity concentrations are expected, the high sensitivity of the instrument could allow for measurements with air sample, with no need for CO$_2$ extraction.
The achieved sensitivity is thus sufficient for applications in nuclear power environment. With its simplicity and robustness, it is the ideal tool for in-situ measurements.
As society is facing the challenge of decommissioning a large number of nuclear power plants worldwide, the work presented here has potential for a great impact.
In the future, we aim to detect even lower level of radiocarbon, of the order of its natural abundance. Very low levels of radiocarbon dioxide could be detected with the instrument and technique presented here, by using long averaging time, and cooling down the sample to reduce interferences from other isotopes.

This work was supported by the EMRP MetroRWM project. The EMRP is funded by the EMRP participating countries within EURAMET and the EU.

\end{document}